\documentclass[twocolumn, aps, prx, superscriptaddress, amsmath, amssymb]{revtex4-2}

\usepackage{graphicx}
\usepackage{dcolumn}
\usepackage{bm}
\usepackage{hyperref}
\usepackage{amsmath}
\usepackage{natbib}
\usepackage[utf8]{inputenc}
\usepackage{float}            
\usepackage{booktabs}
\usepackage{tikz} 

\usepackage[most]{tcolorbox}
\tcbset{colback=gray!10, colframe=gray!60!black, boxrule=0.4pt, arc=2pt, left=6pt, right=6pt, top=4pt, bottom=4pt}

\usepackage{caption} 

\usepackage{subcaption}   

\begin{document}

\title{Emergent evaluation hubs in a decentralizing large language model ecosystem} 

\author{Manuel Cebrian}
\affiliation{Center for Automation and Robotics, Spanish National Research Council, Spain}
\author{Tomomi Kito}
\affiliation{Graduate School of Creative Science and Engineering, Waseda University, Japan}
\author{Raul Castro Fernandez}
\affiliation{Department of Computer Science, University of Chicago, USA}

\begin{abstract}
Large language models are proliferating, and so are the benchmarks that serve as their common yardsticks. We ask how the agglomeration patterns of these two layers compare: do they evolve in tandem or diverge? Drawing on two curated proxies for the ecosystem, the Stanford Foundation-Model Ecosystem Graph and the Evidently AI benchmark registry, we find complementary but contrasting dynamics. Model creation has broadened across countries and organizations and diversified in modality, licensing, and access. Benchmark influence, by contrast, displays centralizing patterns: in the inferred benchmark–author–institution network, the top 15\% of nodes account for over 80\% of high-betweenness paths, three countries produce 83\% of benchmark outputs, and the global Gini for inferred benchmark authority reaches 0.89. An agent-based simulation highlights three mechanisms: higher entry of new benchmarks reduces concentration; rapid inflows can temporarily complicate coordination in evaluation; and stronger penalties against over-fitting have limited effect. Taken together, these results suggest that concentrated benchmark influence functions as coordination infrastructure that supports standardization, comparability, and reproducibility amid rising heterogeneity in model production, while also introducing trade-offs such as path dependence, selective visibility, and diminishing discriminative power as leaderboards saturate.
\end{abstract}

\maketitle
\section{Introduction}
Foundation models (large neural networks pre-trained on web-scale corpora and then fine-tuned for diverse tasks) are central to modern AI. Their footprints vary widely. GPT-4~\cite{openai2023gpt4} is a proprietary-access, multimodal model with public technical documentation and no released weights at the time of writing. BLOOM~\cite{scao2022bloom} is an openly released, 176B-parameter multilingual model from an international consortium with code, weights, and detailed documentation. Baidu’s ERNIE Bot~\cite{baidu2023erniebot} provides public technical information with access via a developer API. These exemplars differ in geography, openness, and modality, reflecting a rapidly diversifying landscape aligned with the machine behavior agenda \cite{Rahwan2019MachineBehaviour}.

Public resources such as the Stanford Ecosystem Graph~\cite{bommasani2023ecosystem} chart this boom, cataloging hundreds of models that differ in size, capability, licensing, transparency, energy footprint, and organizational and geographic origin. For policymakers, developers, and researchers, the breadth of signals to parse (\emph{Who built it? How was it trained? Where can it be used?}) taxes \textit{sensemaking}, the process of turning ambiguity into shared understanding~\cite{Weick1995}.

Benchmarks have become the field’s primary coordination device for evaluation, safety, and societal impact \cite{HernandezOrallo2020Broken,Cebrian2025Authority}.Benchmark creation is geographically and organizationally diverse—spanning open-source collectives, industry labs, and student workshops. Accordingly, we ask whether evaluative attention is diffuse or concentrated.

These observations motivate three research questions:
\begin{itemize}\itemsep0pt
  \item  How have the model–production and benchmark layers co-evolved from 2019–2025?
  \item  Where does inferred benchmark authority concentrate across institutions and countries, and what does that distribution imply for coordination benefits  versus trade-offs?
    \item Which generative mechanisms reproduce the observed heavy-tailed benchmark influence?
\end{itemize}

This pattern parallels cumulative advantage in other knowledge domains, where influence concentrates even as participation broadens \cite{Merton1968Matthew,Price1965Networks,Fortunato2018ScienceOfScience,Wang2013LongTermImpact,Clauset2015Faculty}. As the ecosystem expands, path dependence can reinforce central positions \cite{Barabasi1999Emergence,Salganik2006Inequality,Sinatra2016EvolutionImpact,Wu2019LargeTeams,Liu2018HotStreaks}, yielding heavy-tailed influence with coordination benefits and bounded trade-offs.

\begin{figure*}[!t]
\centering
\begin{subfigure}[t]{0.32\textwidth}
\centering
\includegraphics[width=\linewidth]{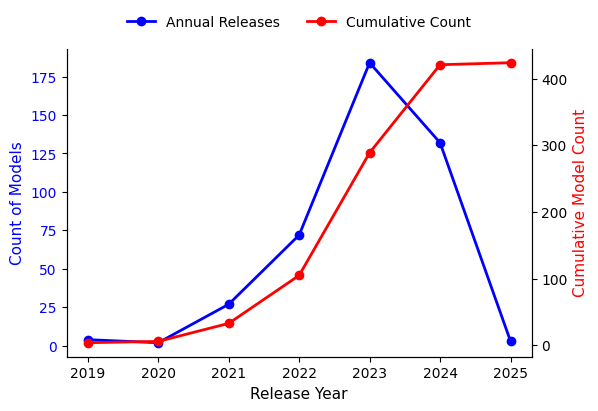}
\caption{} \label{fig:releases}
\end{subfigure}\hfill
\begin{subfigure}[t]{0.32\textwidth}
\centering
\includegraphics[width=\linewidth]{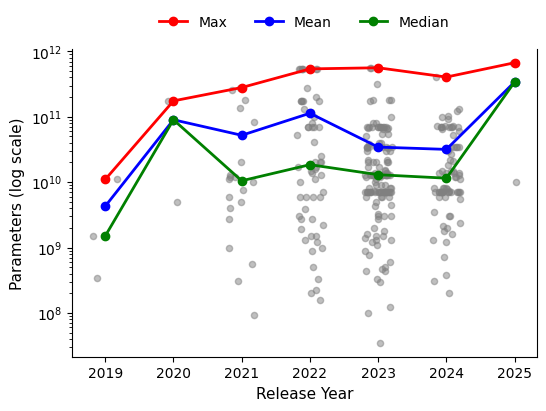}
\caption{} \label{fig:scaling}
\end{subfigure}\hfill
\begin{subfigure}[t]{0.32\textwidth}
\centering
\includegraphics[width=\linewidth]{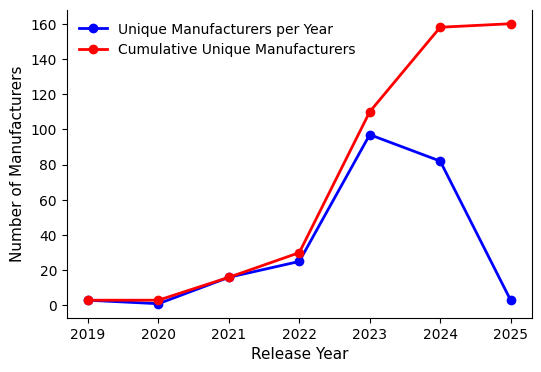}
\caption{} \label{fig:producers}
\end{subfigure}
\caption{Growth of the foundation-model ecosystem.
(a) Annual and cumulative model releases, 2019–early 2025 (2025 is partial-year).
(b) Reported parameter counts (log scale), 2019–2025.
(c) New and cumulative manufacturers per year; over 160 organizations by early 2025.}
\end{figure*}

We study these questions using two high-quality datasets: the Stanford Ecosystem Graph for models and the Evidently AI repository for benchmarks. We measure (i) the tempo and diversification of model releases (counts, modalities, documentation), (ii) the parallel expansion of benchmarks (volume, citation velocity, open-source engagement), and (iii) whether the emerging benchmark network reflects broader community governance or concentrated influence with coordination benefits.

We combine interpretable metrics, network analysis of inferred benchmark–author–institution links, and an agent-based simulation with three policy levers (entry rate $\gamma$, reuse friction $\beta$, adoption responsiveness $\delta$) to map where evaluative attention concentrates and how it can shift. Conceptually, we build on science-of-science accounts of cumulative advantage and heavy-tailed attention, collaboration structure, and integrative syntheses \cite{Merton1968Matthew,Price1965Networks,Barabasi1999Emergence,Clauset2009Power,Newman2001Collab,Fortunato2018ScienceOfScience,Evans2011Metaknowledge,WangBarabasi2021Book, Wang2013LongTermImpact,Sinatra2016EvolutionImpact,Liu2018HotStreaks,Ke2015SleepingBeauties,Wuchty2007Teams,Uzzi2013Atypical,Foster2015Novelty,Fleming2001Recombinant,Wu2019LargeTeams,Clauset2015Faculty}.

Rather than speculate from anecdotes, we offer a reproducible, computational-social-science account of how AI development and evaluation co-evolve. Using network-science tools and citation/usage–based influence measures \cite{wang2021science}, we map where evaluative attention and coordination accrue, quantify the degree and dynamics of (de)concentration, and surface conditions under which influence shifts. This structural lens clarifies who effectively sets the yardsticks of “success” and when, providing an auditable basis for researchers and policymakers to reason about governance, transparency, and the design of evaluation infrastructures in a rapidly changing field.

\section{Datasets}
Our analysis draws on two complementary datasets that, together, capture both the supply side (models) and the evaluation side (benchmarks) of the foundation-model landscape, enabling comparisons across layers of the ecosystem.

The first dataset is the Stanford Foundation-Model Ecosystem Graph (snapshot 2025-03-01) \cite{bommasani2023ecosystem}. A monthly crawler aggregates releases mentioned in arXiv preprints, model cards, Hugging Face pages, GitHub tags, and company blogs, merges aliases and checkpoints, and verifies external links. After removing seventeen records with ambiguous launch dates we retain 418 distinct models released between January 2018 and 28 February 2025. For every model we keep its release date, licence class, declared modalities, and any reported parameter count. From these fields we derive three analysis variables: (i) the number of supported modalities; (ii) a binary “full documentation” flag set when a model card, a training-data summary, and a licence text are all present; and (iii) the publisher’s region, assigned from a hand-curated headquarters table with an API fallback for missing cases. Coverage is broad but metadata depth is uneven, so our indicators mildly favour well-documented releases.

Our regostru contains 248 unique LLM benchmarks and evaluation datasets \cite{EvidentlyLLMBenchmarks2025}. Applying our inclusion criteria—public paper, code, and data under a permissive/open license—yields 134 eligible suites spanning capabilities and safety (including bias/toxicity).”

The second dataset is the Evidently AI open registry of LLM benchmarks \cite{EvidentlyLLMBenchmarks2025}, snapshot 12 June 2025 , which lists 248 benchmark suites spanning language understanding, reasoning, safety, code generation, retrieval-augmented generation, and multimodal tasks. Inclusion requires that data, code, and methodological write-ups be public under a permissive licence, excluding opaque suites. For each benchmark we extract its arXiv identifier, pull the full author roster and both total and monthly citation counts from the Semantic Scholar API, and scrape GitHub engagement statistics (stars, forks, watchers, open-issue counts, and last-push date) via the official REST API. Sample sizes are parsed to numeric counts. To infer institutional affiliations, we issue one LLM query per \emph{(paper, author)}: for each author we retrieve the paper title and publication year from arXiv, prompt an LLM (Gemini 2.5 Flash) to return a single line in the format “\textit{Institution, Country}” representing the author’s primary affiliation at that year, take the first line of the reply, and split on the last comma to parse institution and country; we then aggregate these per paper. We do not perform alias resolution or ROR/GRID mapping, and countries are taken verbatim, so temporal or naming inconsistencies may remain (details in Methods). All usage signals are collected on the same day to minimise timing bias, repository commit histories are preserved so analyses can be tied to exact tags, and a log-scaled “authority” index is computed by blending citations, GitHub engagement, sample size, and team size. The dataset is therefore audit-ready and longitudinally consistent, albeit selective and dependent on heuristic affiliation resolution.

Taken together, the model graph and benchmark registry provide a time-stamped, quality-controlled view of which models enter the field, who releases them, how completely they are documented, and which tests the community deploys to measure their capabilities. Though lean, the pair is high-quality by design: public, versioned, machine-readable sources with strict inclusion (paper + code + data under a permissive license), de-duplication, and stable IDs that enable auditable linkages and longitudinal analyses—favoring fidelity over coverage. These paired sources form the empirical foundation for all structural analyses that follow and supply shared signals for sensemaking across an increasingly heterogeneous ecosystem.

\section{Results: Evolution of the Model Ecosystem}

As illustrated in Fig.\ref{fig:releases}, foundation-model output was essentially flat through 2020, rose modestly in 2021, and then entered an accelerating phase: annual releases tripled in 2022 and exceeded 180 in 2023, pushing the cumulative total above 400 by early 2025. As shown in Fig.\ref{fig:scaling}, the scale of foundation models ballooned in 2020 and has since maintained a frontier near the trillion-parameter mark, while the median model continues to creep upward. This indicates that extreme-scale models have not yet displaced a long-tail population of smaller models.

As shown in Fig.~\ref{fig:producers}, the supply side of the ecosystem has shifted from a handful of well-known labs to a broad, decentralized field. No more than two new model producers appeared in any year before 2021; by contrast, 2023 alone added ninety-five first-time manufacturers and pushed the cumulative total of distinct model publishers above 110. A further wave in 2024 lifted the running count to more than 150 organizations. This diversification increases coordination demands alongside model complexity, as a rapidly widening array of corporate, academic, and open-source actors contributes to the field.

\begin{figure*}[!ht]
\centering
\begin{subfigure}[t]{0.64\textwidth}
\centering
\includegraphics[width=\linewidth]{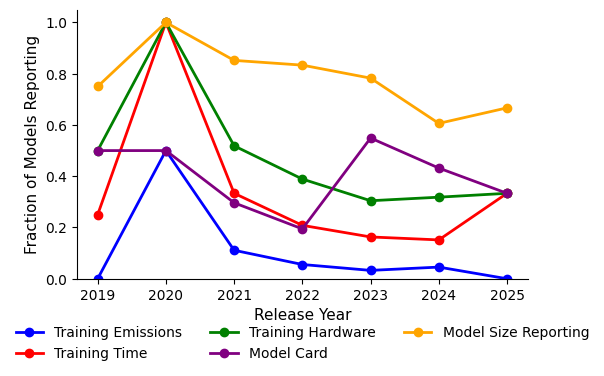}
\caption{} \label{fig:transparency}
\end{subfigure}\hfill
\begin{subfigure}[t]{0.34\textwidth}
\centering
\includegraphics[width=\linewidth]{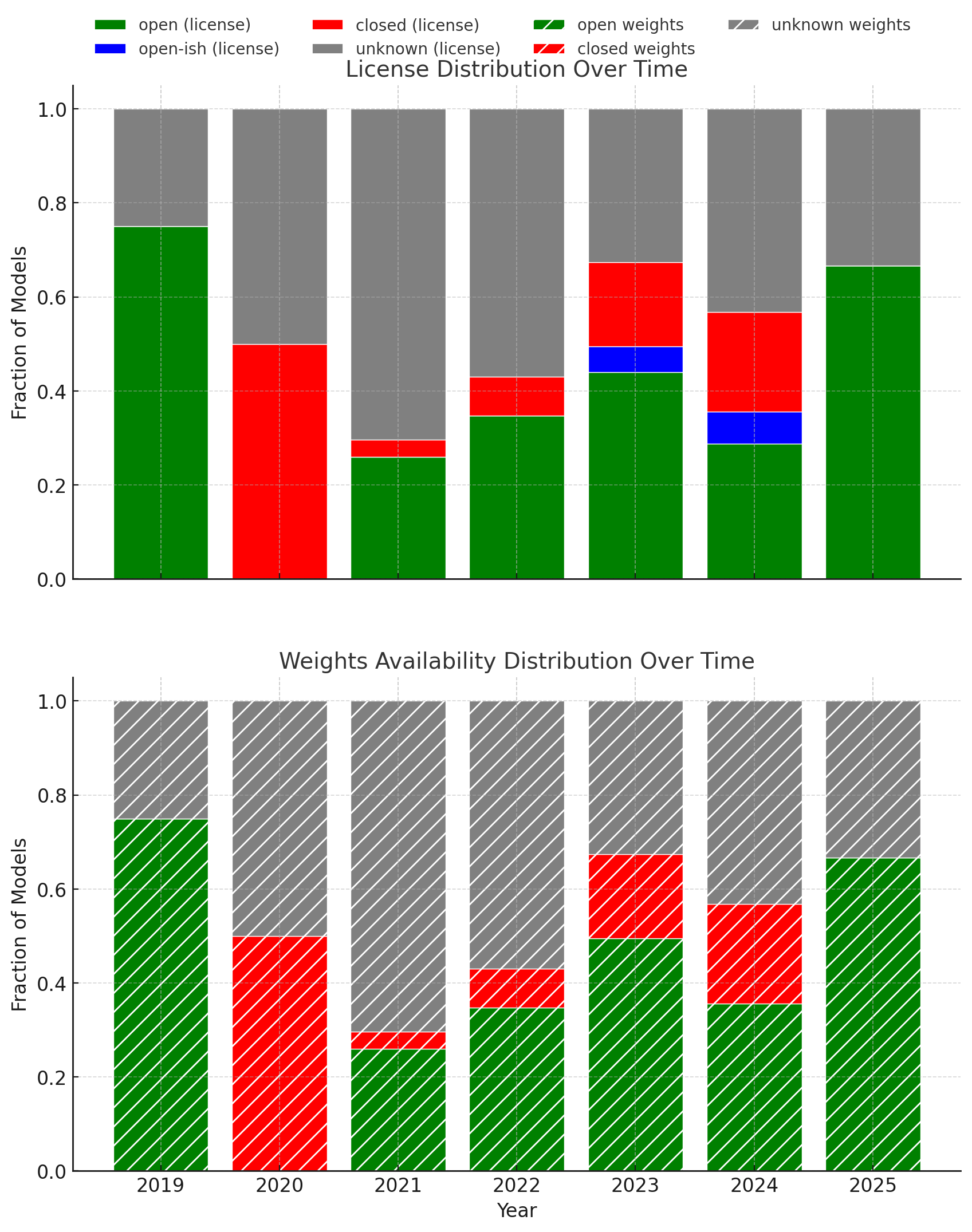}
\caption{} \label{fig:access}
\end{subfigure}

\caption[Documentation and access trends, 2019–2025]{%
Documentation and access trends, 2019–2025.
(a) Fraction of new models disclosing training emissions (blue), training time (red), training hardware (green), structured model cards (purple), and explicit parameter counts (orange). All metrics peak in 2020 then decline; size reporting remains at $\approx 60\%$ by 2024.\
(b) Access conditions for foundation models, 2019–2025. Top panel: License status of newly released models, binned as permissive open source (green), partially open or “community” licenses such as LLaMA~2 (blue), fully closed licenses (red), and cases where the license is not disclosed (gray). Bottom panel: Availability of pre-trained weights, recorded as openly downloadable (green), gated or paywalled (red), or unspecified (gray). The share of fully open licenses and weights plummets after 2019, bottoms out in 2021, and then recovers only partially—never exceeding 45–50\% of annual releases. Closed or ambiguous terms remain common, indicating that rapid ecosystem growth has not been matched by equivalent gains in access transparency.
}
\end{figure*}

Figure~\ref{fig:transparency} reveals that transparency has not kept pace with the accelerating output of foundation models. A short-lived high-water mark in 2020, driven by a few unusually well-documented flagship releases, gave way to a steady erosion: explicit reporting of training emissions, hardware, and runtime now appears only sporadically, and even basic metrics like parameter counts are omitted in roughly two-fifths of new models. The brief rebound of formal model cards in 2023 suggests growing community awareness, yet overall the data imply that documentation quality is inversely correlated with the speed at which new models enter the ecosystem.

Figure~\ref{fig:access} highlights a persistent tension between rapid model proliferation and open access. Whereas three-quarters of 2019 releases shipped with permissive open-source licenses and downloadable weights, that fraction collapsed during the 2020–2021 surge, when closed or unspecified terms became the norm. A partial rebound in 2023 coincides with high-profile “community” licenses (e.g., LLaMA~2’s license) but still leaves roughly half of new models either fully closed or ambiguous with respect to usage rights. The pattern is mirrored in weight availability, underscoring that license text and practical access typically move in lockstep. This fragmented landscape increases the value of shared yardsticks for independent evaluation and reuse across heterogeneous access regimes.

Table \ref{tab:geo} confirms an uneven geography in our sample: roughly half of all documented foundation models originate from the United States, with China and the United Kingdom comprising the next two largest contributors. A long tail of other countries accounts for fewer than ten models each, while 79 releases list no verifiable headquarters location (``Unknown''), underscoring the limits of publicly available provenance data. Overall, activity is concentrated in US institutions with notable hubs in China and the UK; large regions of Africa and South America remain essentially absent from the current foundation-model ecosystem.

\begin{table}[!ht] \centering \caption{Foundation-model releases by country of the publisher’s headquarters (2019–2025 snapshot).} \label{tab:geo} \begin{tabular}{lc} \hline Country & Number of Models \\ \hline United States of America & 214 \\ Unknown / Not disclosed & 79 \\ China & 50 \\ United Kingdom & 39 \\ Canada & 12 \\ South Korea & 8 \\ France & 7 \\ Israel & 6 \\ Germany & 5 \\ Singapore & 2 \\ United Arab Emirates & 2 \\ Japan & 1 \\ Russia & 1 \\ Spain & 1 \\ \hline \end{tabular} \end{table}

\begin{figure*}[!t]
\centering
\begin{subfigure}[t]{0.5\textwidth}
\centering
\includegraphics[width=\linewidth]{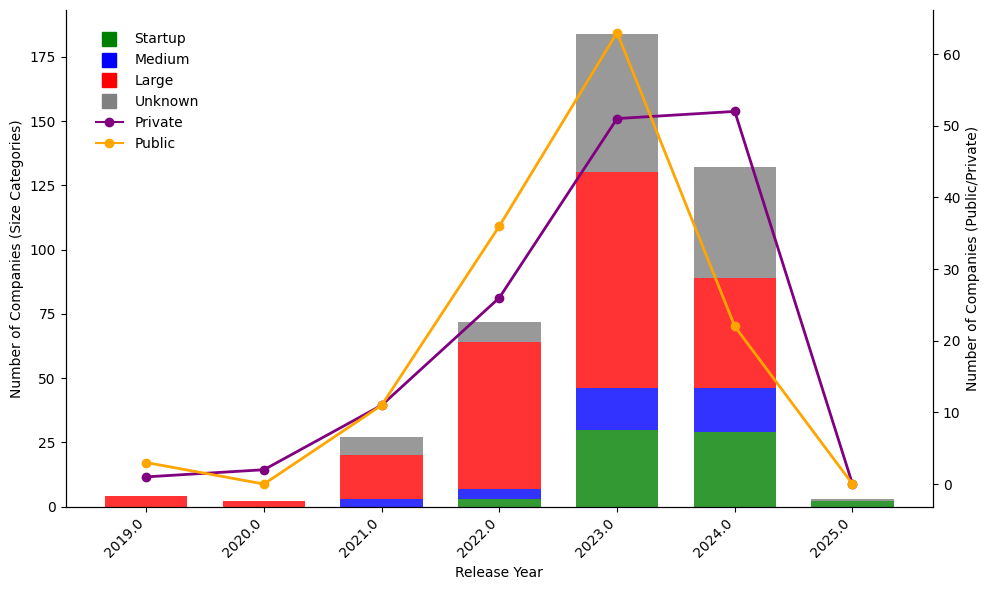}
\caption{} \label{fig:corporate}
\end{subfigure}\hfill
\begin{subfigure}[t]{0.5\textwidth}
\centering
\includegraphics[width=\linewidth]{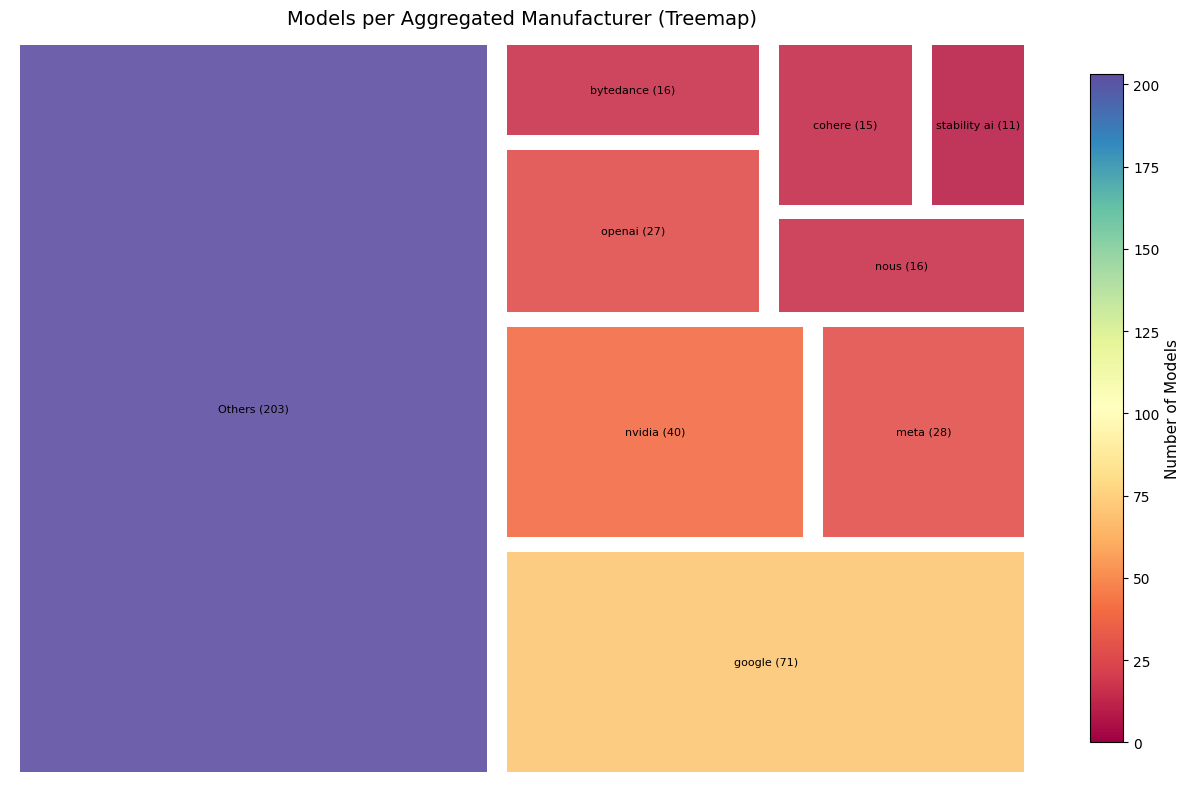}
\caption{} \label{fig:longtail}
\end{subfigure}

\caption[Corporate footprint and concentration patterns]{%
Corporate footprint and concentration patterns, 2019–2025.
(a) \emph{Shifting corporate footprint:} stacked bars (left axis) show, by release year, the number of producing organisations classified as start-up (green), medium-sized (blue), large (red), or unknown (gray); superimposed lines (right axis) plot private (purple) and publicly traded (orange) entrants. Pre-2021 activity is negligible and driven by large public firms. The ecosystem broadens in 2022 and peaks in 2023 with over 180 distinct companies ($\approx  ~30$ start-ups). In 2024 total producers dip modestly while private entrants keep rising and public-company entries fall sharply. (2025 is partial-year.)
(b) \emph{Concentration and long tail:} treemap area (and color shading) is proportional to the number of distinct models per aggregated organisation. “Others” groups 203 models across more than 100 smaller actors.
}
\end{figure*}

\begin{figure}[!h]
\centering
\includegraphics[width=0.4\textwidth]{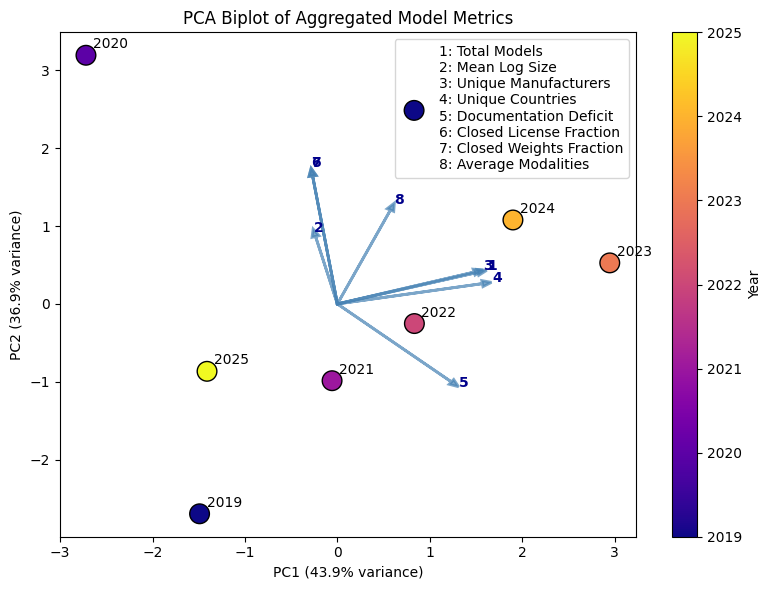}
\caption{We applied PCA to eight yearly $z$-scored metrics; the top two components explain about 81\% of the variance.
PC1 reflects overall growth—more models, larger size, and more manufacturers and countries.
PC2 reflects openness, with higher values for more modalities and lower ones for poor documentation and closed weights.
}
\label{fig:pca}
\end{figure}

Figure~\ref{fig:corporate} shows that the recent boom in foundation models is no longer confined to a small set of well-capitalized public tech giants. Large corporations remain the single biggest slice of activity, but their relative share decreased after 2022 as start-ups and medium-sized firms crowded in. The parallel rise of privately held entities—and an abrupt drop in new publicly traded entrants during 2024—suggest a financing pivot from listed companies toward venture-funded or privately backed labs. This shift further diversifies incentives and oversight approaches, as different classes of organizations (big tech, startups, academia, etc.) may face distinct governance challenges.

Figure~\ref{fig:longtail} illustrates a dual reality of the model ecosystem: a handful of hyperscale labs account for a large share of headline output, yet nearly half of all models originate from a diffuse population of smaller or single-release organizations. For example, the main manufacturer of models alone accounts for about 17\% of the total model count in our sample, and the combined share of the next seven most prolific producers reaches roughly 52\%. The remaining 200+ models are produced by more than one hundred distinct companies, underscoring the increasingly decentralized nature of foundation-model development and the associated coordination demands.

Univariate trends make clear that “everything” is rising—model counts, producer countries, organizational diversity—while documentation quality and weight availability lag behind. What is less obvious is how these dimensions interact: do years with explosive scale also suffer larger transparency gaps, or are the trends independent? To answer this, we apply a principal component analysis (PCA) that projects eight annual ecosystem indicators onto two orthogonal axes capturing over 80\% of total variance (Fig.~\ref{fig:pca}). The first principal component (PC1) bundles the expansion signals (total model count, mean log-parameters, number of unique manufacturers, number of countries) and can be interpreted as a general \textit{expansion} axis. The second component (PC2) contrasts openness with opacity: it scores high when documentation completeness and open weights are common, and low when those are deficient or closed, effectively capturing a \textit{transparency} axis. Annual markers move steadily away from the origin on both axes, showing that the ecosystem is becoming simultaneously larger \textit{and} more uneven in information quality. In other words, the effort a stakeholder must expend to make sense of the landscape grows every year. PCA thus condenses a tangle of separate trend lines into a single visual synopsis whose geometry makes the conclusion unmistakable: rapid quantitative growth in foundation models has been accompanied by a multi-dimensional broadening of the governance burden, including partial recoveries and relapses in openness.

\section{Benchmark Expansion and Centralization}

\begin{figure*}[!t]
\centering
\begin{subfigure}[t]{0.48\textwidth}
\centering
\includegraphics[width=\linewidth]{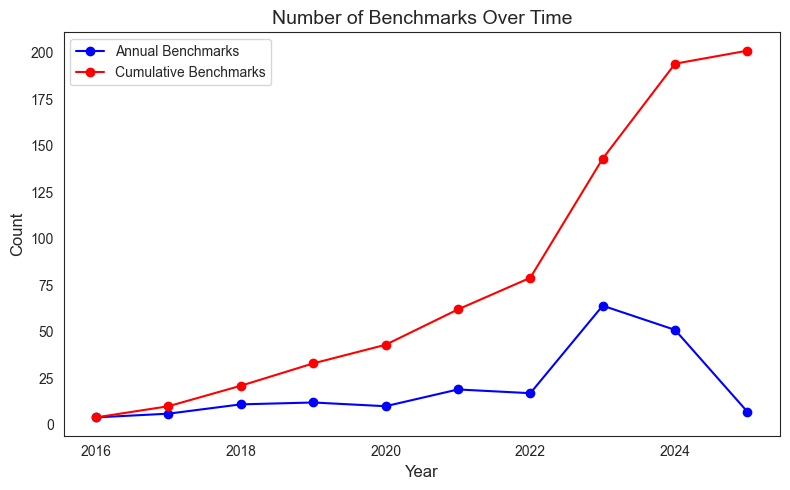}
\caption{} \label{fig:benchCount}
\end{subfigure}\hfill
\begin{subfigure}[t]{0.48\textwidth}
\centering
\includegraphics[width=\linewidth]{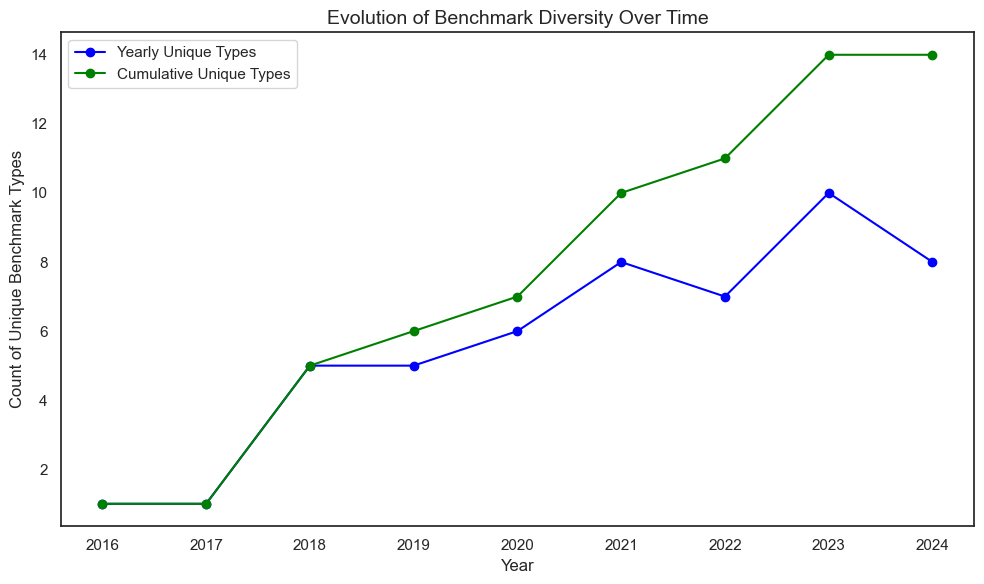}
\caption{} \label{fig:benchmark_diversity}
\end{subfigure}

\medskip
\begin{subfigure}[t]{0.48\textwidth}
\centering
\includegraphics[width=\linewidth]{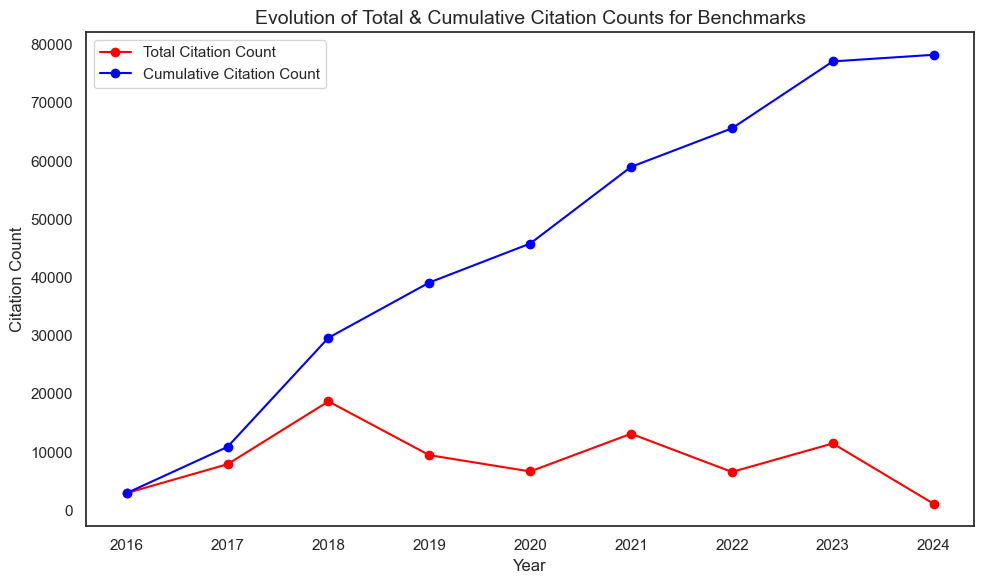}
\caption{} \label{fig:benchmark_citations}
\end{subfigure}\hfill
\begin{subfigure}[t]{0.48\textwidth}
\centering
\includegraphics[width=\linewidth]{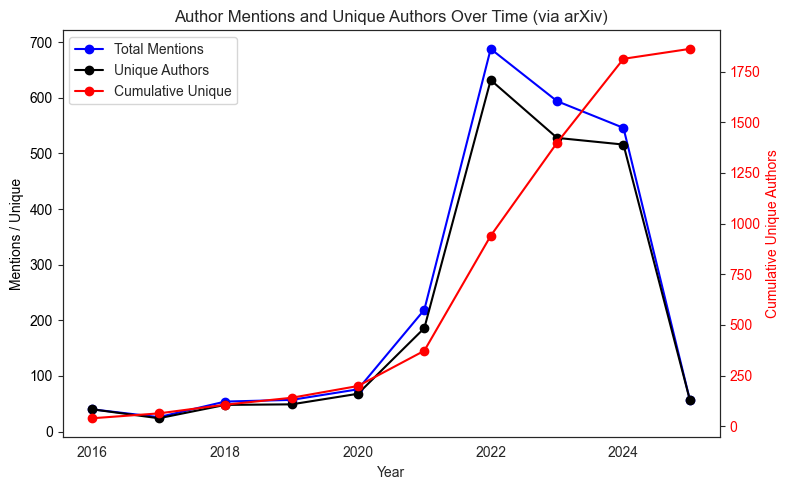}
\caption{} \label{fig:benchAuthors}
\end{subfigure}

\caption[Benchmark-ecosystem growth]{%
Benchmark-ecosystem growth, 2016–2025.
(a) Annual releases and cumulative stock surpass 100 benchmark suites by 2024.
(b) Benchmark categories widen from one in 2016 to fourteen by 2024, with five new types in 2023 alone.
(c) Citations top 75,000 in 2024, spiking around landmark suites in 2018, 2021, and 2023.
(d) Author participation accelerates after 2020—both total and unique contributors—pushing the cumulative author pool sharply upward.
}
\end{figure*}

The second half of our analysis turns from model development to the state of evaluation. Using the Evidently AI registry as a high-precision sample of public benchmarks, we document a sharp post-2021 acceleration in benchmark introductions, with a pronounced surge in 2023 and sustained, elevated activity through 2025 (Fig. \ref{fig:benchCount}). To cross-check against the broader literature stream, we train a lightweight text classifier (TF--IDF over title+abstract with $\ell_2$-regularized logistic regression) on Evidently positives versus randomly sampled non-benchmark LLM papers and apply it to our monthly arXiv crawl. Figure~\ref{fig:benchCount} reports monthly counts at two probability thresholds ($\ge 0.5$ likely,'' $\ge 0.8$ very likely'') alongside all LLM papers; the inset plots a score-weighted volume (monthly sum of predicted probabilities of being a benchmark according to our model). Across both datasets, the headline result is growth: benchmark activity has shifted from sporadic to sustained, rising far above pre-2021 levels and remaining high thereafter. We use the arXiv view as a sanity check; all structural analyses rely on the curated Evidently set.

\begin{figure}[t]
\centering
\begin{tikzpicture}
\node[anchor=south west, inner sep=0] (main) at (0,0)
{\includegraphics[width=\linewidth]{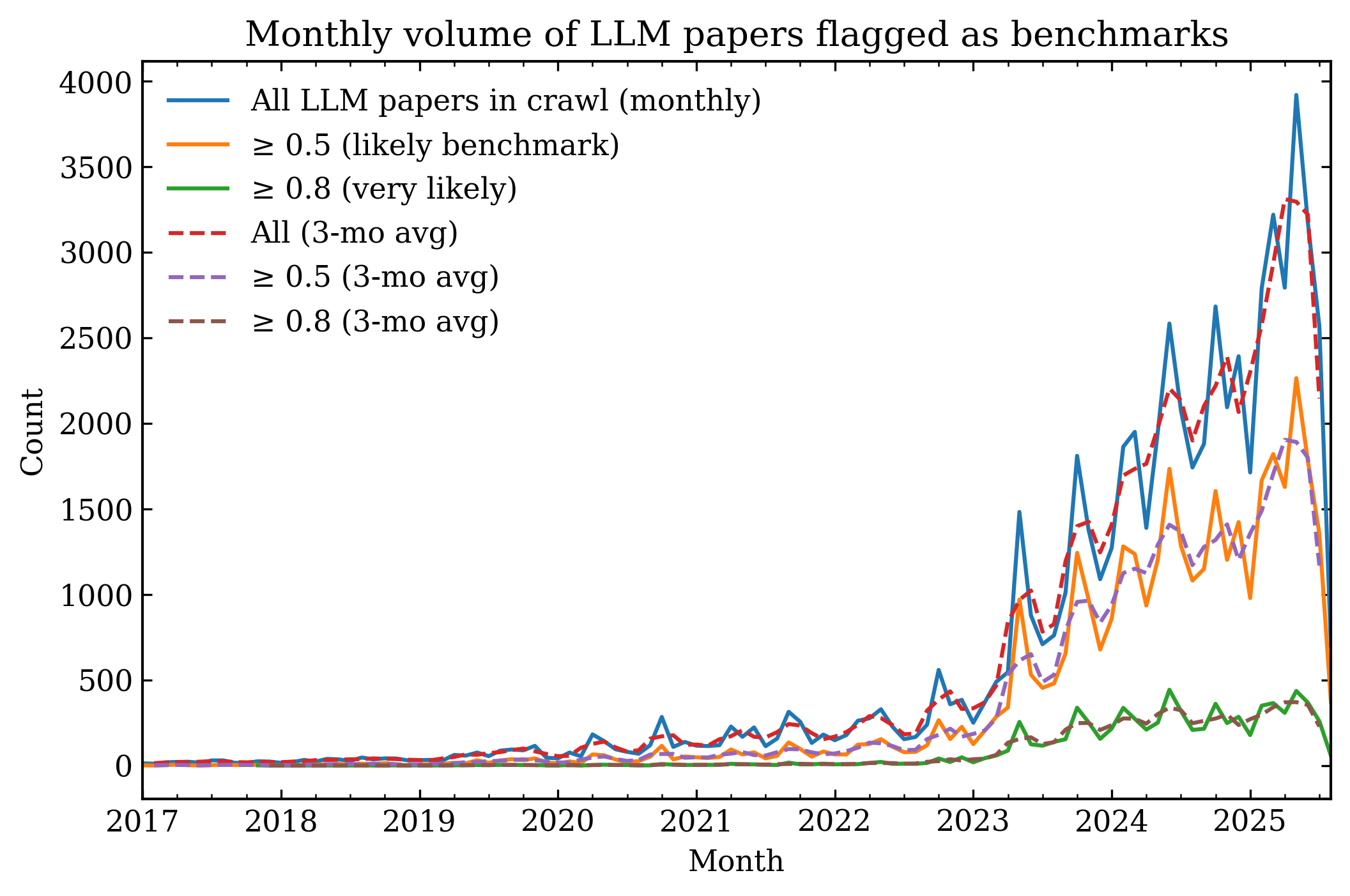}};
\node[anchor=north east, inner sep=0] at ([xshift=-120pt,yshift=-70pt]main.north east)
{\includegraphics[width=.40\linewidth]{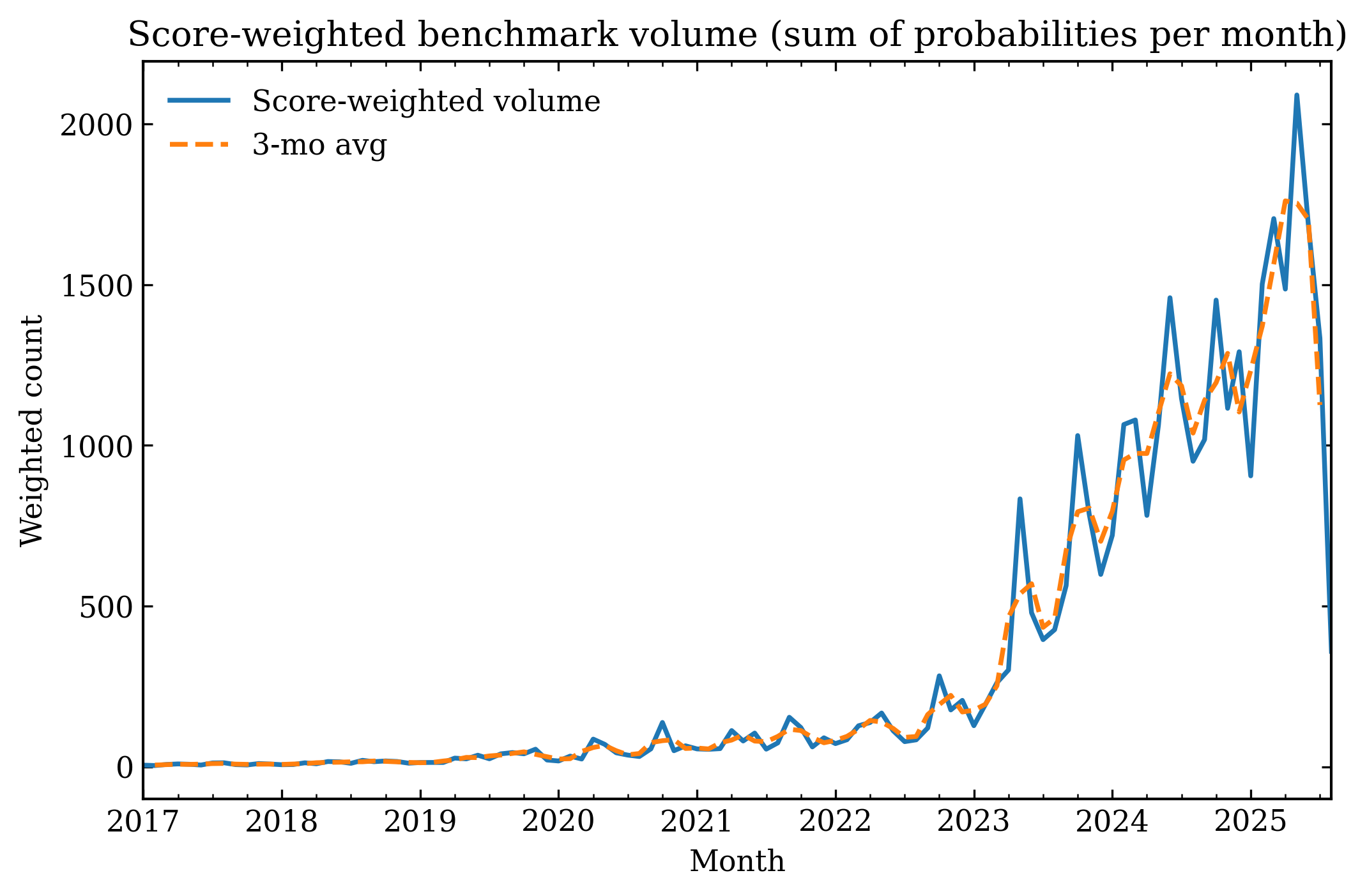}};
\end{tikzpicture}
\caption{Benchmark growth signals. Main panel: monthly counts of arXiv LLM papers flagged by our classifier as likely benchmarks at two thresholds (0.5, 0.8), with 3-month moving averages; solid blue shows all LLM papers in the crawl. Inset: score-weighted benchmark volume (sum of predicted probabilities per month) with 3-month average, which smooths threshold effects.}
\label{fig:benchCount}
\end{figure}

In addition to sheer quantity, benchmark content has become increasingly specialized and diverse (Figure \ref{fig:benchmark_diversity}). Recent benchmarks target a wide range of model capabilities and domains, including core language understanding, logical reasoning, code generation, factual retrieval with external knowledge, safety and bias assessment, and multimodal (e.g., vision-and-language) tasks. This diversification in benchmark scope reflects a broadening of the community’s evaluative focus to match the multifaceted challenges posed by new models.

Benchmark \textit{authorship} has scaled even more steeply than the benchmarks themselves. Figure~\ref{fig:benchAuthors} shows that annual author mentions in benchmark papers remained below 100 until 2020, then jumped to 219 in 2021 and peaked at 688 in 2022. Correspondingly, the pool of distinct contributors grew from only about 40 individuals in 2016 to more than 600 in 2022, while the cumulative count of unique benchmark authors climbed past 1,800 by early 2025. This influx expands the set of perspectives informing evaluation and, in parallel, increases coordination demands as the contributor base becomes more decentralized.

\begin{figure}[h!]
\centering
\includegraphics[width=\linewidth]{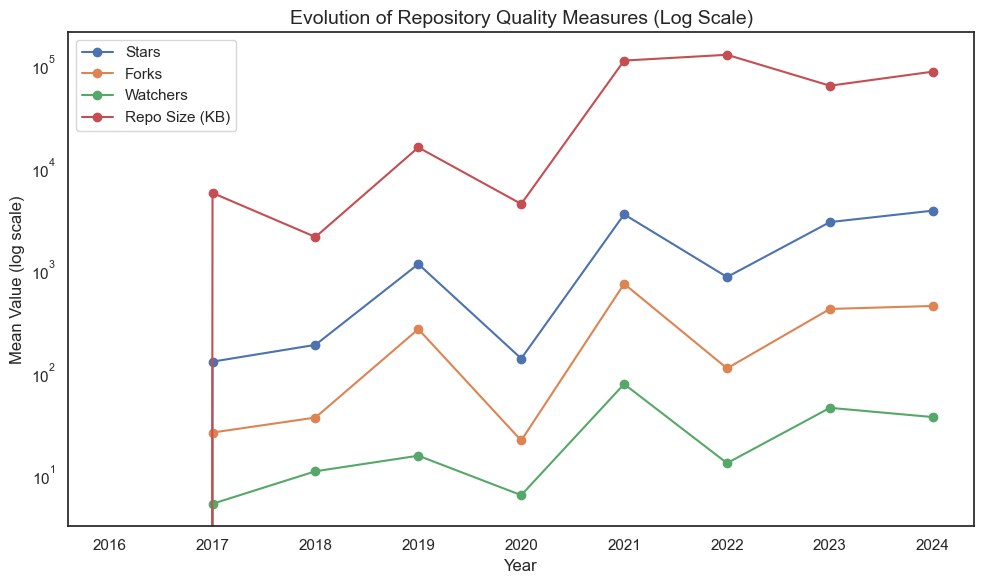}
\caption{Mean GitHub stars, forks, watchers, and repository size are plotted on a log scale.
The pronounced post-2020 uptick—especially in stars and forks—signals accelerated
community uptake of evaluation suites, while the surge in repository size reflects
richer supporting assets (e.g., larger datasets, interactive dashboards).}
\label{fig:repo_quality}
\end{figure}

Beyond raw counts, newer benchmarks also exhibit heightened impact as measured by citation and engagement metrics. As can be observed in Figure \ref{fig:benchmark_citations}, the average benchmark introduced after 2021 accrues citations at a higher monthly rate than those from earlier years, indicating that recent evaluation suites are being picked up and referenced in the literature more quickly. Likewise, as can be seen in Figure \ref{fig:repo_quality}, many benchmarks released with open-source code are seeing substantial developer engagement: it is now common for a benchmark’s repository to garner hundreds or even thousands of GitHub stars within its first year. For instance, the introduction of open evaluation platforms for chat-based LLMs in 2023 (e.g., multi-task chatbot ``arena'' benchmarks) attracted tens of thousands of users and quickly became reference points for comparing dialogue models. Such community enthusiasm, reflected in both academic citations and open-source contributions, underscores the growing centrality of benchmarking in the LLM research ecosystem.

\begin{figure*}[htbp]
\centering
\includegraphics[width=\textwidth]{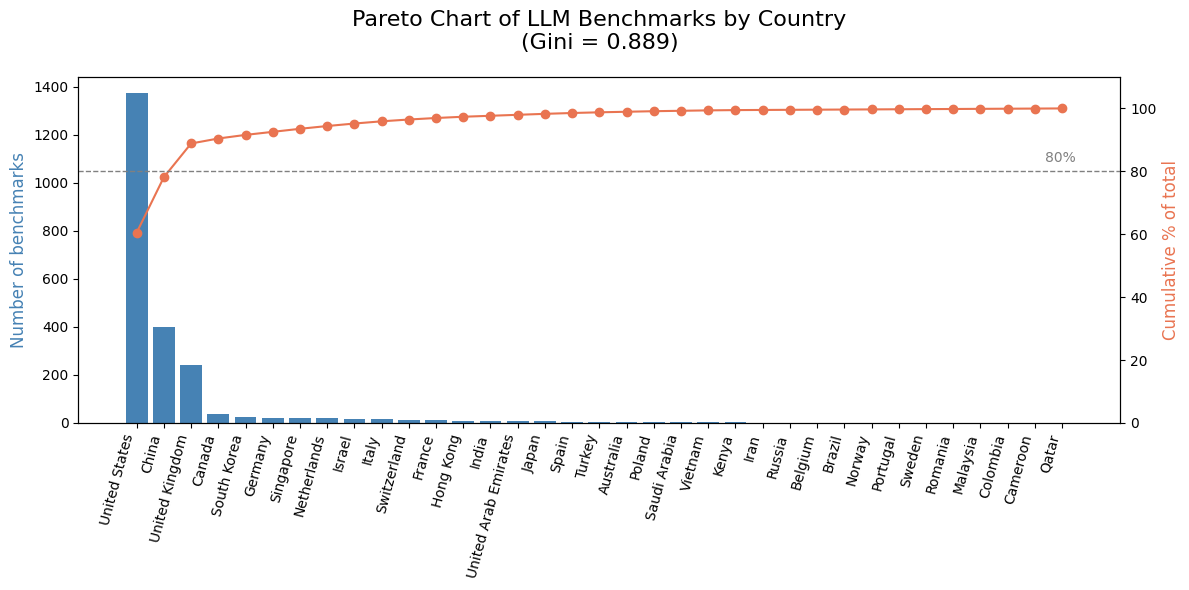}
\caption{%
Pareto chart of benchmark origin by country inferred from paper-year author affiliations. Blue bars (left axis) count
benchmarks; the orange line (right axis) shows cumulative share. The
United States, China, and the United Kingdom together exceed the
80\% threshold (grey line), yielding a Gini coefficient of 0.889. Below
Canada the drop is steep: no other nation tops 25 benchmarks, and a
twenty-country tail contributes under 10\%. The head–tail split reveals
a highly uneven global footprint despite rapid ecosystem growth. Country counts reflect inferred paper-year affiliations within our curated sources and likely underrepresent smaller and non-English initiatives.}
\label{fig:benchmark_countries}
\end{figure*}

\begin{figure*}[!ht]
\centering
\includegraphics[width=\textwidth]{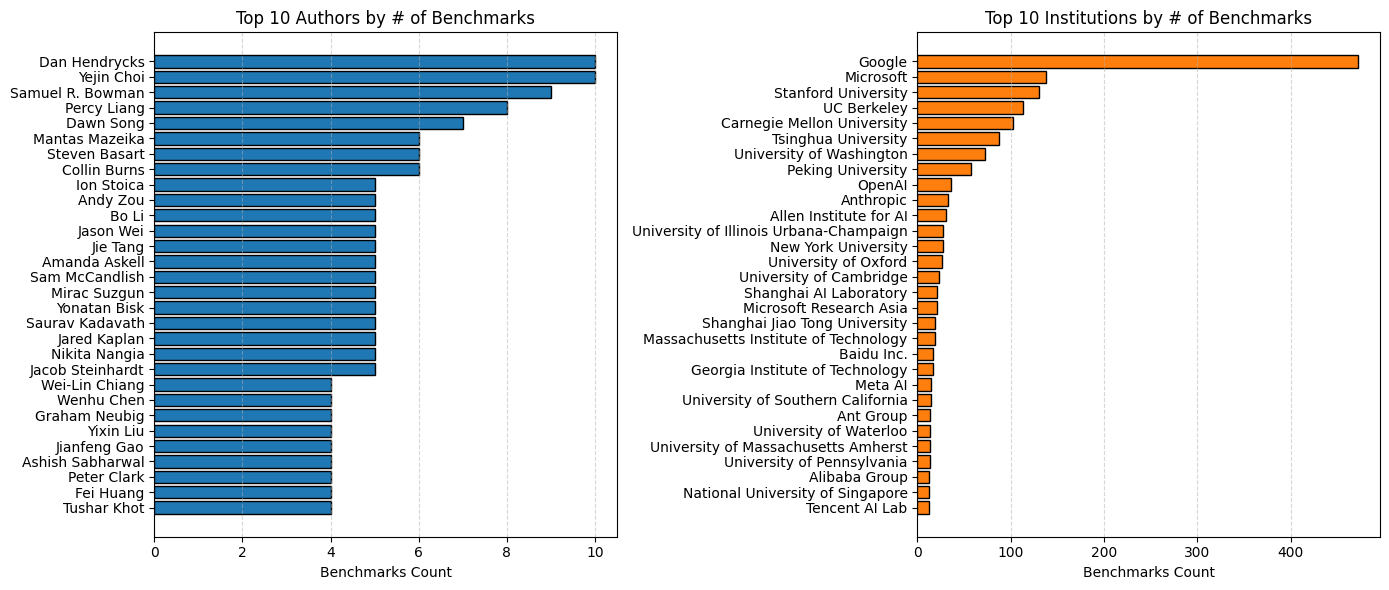}
\caption{%
\emph{Left panel}: the most prolific individual contributors, measured by the number of benchmark suites on which
they appear as an author or co-author. \emph{Right panel}: the institutional leaders, ranked by the total number of
benchmarks that list at least one affiliated author. More than 1,800 unique researchers from hundreds of
organisations have participated in benchmark creation (cf.\ Fig.~\ref{fig:benchAuthors}), indicating broad engagement; at the same time, output follows a heavy-tailed pattern: a handful of researchers contribute to six or more benchmarks, and a small group of tech firms and elite
universities collectively account for the largest share of
high-impact benchmarks. This pattern highlights how central actors can provide shared reference points for
standardization and comparability even as overall community participation expands.}
\label{fig:benchPower}
\end{figure*}

Alongside this broad-based growth in participation, we observe a persistent concentration of measured evaluative influence among a small cluster of organizations (Figure \ref{fig:benchPower}) and countries (Figure \ref{fig:benchmark_countries}) based on inferred affiliations. In our snapshot, this concentrated influence provides widely recognized reference points for comparison while carrying familiar trade-offs such as path dependence and over-optimization risks.

To move beyond anecdote, we define a continuous \emph{benchmark–authority} score that integrates both scholarly attention and developer uptake.
For every benchmark $b$ we compute an influence weight
\[
a_b \;=\; \log\!\bigl(1 + c_b\bigr)
           \;+\;
           \alpha\,\log\!\bigl(1 + s_b\bigr),
\qquad
\alpha = 0.25
,
\]
where $c_b$ is the benchmark’s citation count and $s_b$ the number of GitHub \textit{stars}.  
The logarithm dampens heavy-tailed counts, while the scaling factor $\alpha$ places lesser—but non-negligible—emphasis on open-source engagement relative to citations \cite{Wang2023SoLo}.
Authority is then allocated fractionally across the $n_b$ distinct institutional affiliations associated with the benchmark paper or data-card: an institution $i$ receives
\[
A_i \;=\;
\sum_{b \in \mathcal{B}_i} \frac{a_b}{n_b},
\]
where $\mathcal{B}_i$ is the set of benchmarks with at least one author from institution~$i$.
In effect, $A_i$ aggregates the logarithmically scaled \emph{impact} of all benchmarks linked to $i$, weighted by that institution’s share of authorship credit.

\begin{figure*}[!ht]
\centering
\includegraphics[width=\textwidth]{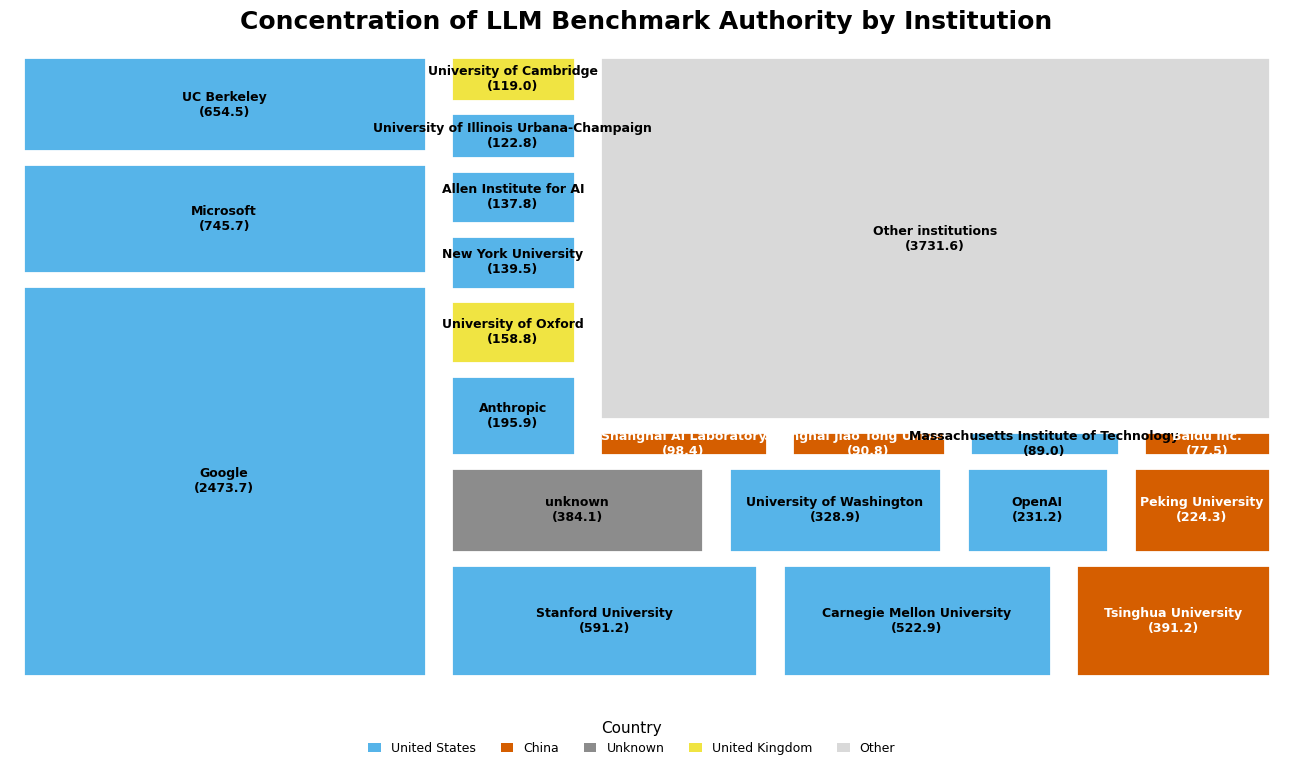}
\caption{Concentration of LLM benchmark authority by institution. Treemap areas are proportional to each institution’s log-scaled authority score, computed from citation and GitHub engagement metrics. The top three institutions collectively hold nearly 50\% of measured benchmark authority in our snapshot, reflecting a heavy-tailed pattern in which central actors provide widely used reference points for standardization and comparability. Names are displayed for auditability; inclusion implies neither endorsement nor ranking. ‘Unknown/unlisted’ indicates affiliations not reliably extracted.}
\label{fig:concentration-authority}
\end{figure*}

The resulting distribution of $A_i$ is highly concentrated \footnote{We down-weight stars because they live on a larger, more volatile scale than citations; \(\alpha=0.25\) was chosen ex ante for conservatism. Results are robust: for \(\alpha\in\{0,0.25,0.5\}\), institution ranks are unchanged (Spearman \(=1.0\)), the top-10 set is identical (Jaccard \(=1.0\)), and concentration (HHI) varies by \(<10^{-4}\). Z-score standardizing each signal yields the same ordering; we retain the log-sum with \(\alpha\) for interpretability and clean ablations (\(\alpha=0\) \(\equiv\) citations-only).}.
Let $\mathcal{I}$ denote the set of all 424 unique institutions in our sample and let $A_{\text{tot}}=\sum_{j\in\mathcal{I}}A_j$ be the total authority mass.
The share held by institution $i$ is $\sigma_i = A_i/A_{\text{tot}}$.
We find
\[
\sigma_{(1)} \approx 0.31,
\quad
\sigma_{(2)} \approx 0.094,
\quad
\sigma_{(3)} \approx 0.083,
\]
so that
\[
\sigma_{(1)} + \sigma_{(2)} + \sigma_{(3)}
\;\gtrsim\; 0.49,
\]
meaning the top three entities alone account for nearly one-half of all benchmark authority (Figure \ref{fig:concentration-authority}).
Extending to the top ten organisations raises the cumulative share above 60\%.
By contrast, the bottom 300+ institutions each account for less than 0.1\,\% of the total.
The concentration is also evident in the Gini coefficient of the authority distribution, $G \approx 0.89$, which indicates unusually high concentration for scientific artifacts \cite{Liu2023EditorialBoards}.

Notably, this evaluative concentration far exceeds the inequality in model production itself. For example, the most prolific model producer accounts for only about 17\% of all models in our dataset—a large share, but nowhere near the $\approx 31$ benchmark authority we find for the top benchmark contributor. In other words, measured influence over evaluation appears more concentrated than measured influence over model development in our sample.

We recomputed authority under $\alpha\in\{0,\,0.25,\,0.5\}$ and compared (i) Top–10 membership via the Jaccard similarity and (ii) Top–20 ordering via Spearman rank computed over the union of entities. For every $\alpha$ pair, both metrics equal $1.0$, indicating exact invariance of the top set and its ordering. Institutional concentration (HHI) changes monotonically but trivially from $\mathrm{HHI}(\alpha{=}0)=0.04200946$ to $\mathrm{HHI}(\alpha{=}0.25)=0.04200146$ and $\mathrm{HHI}(\alpha{=}0.5)=0.04199466$; the absolute change is $\Delta=1.48\times10^{-5}$ (a $-0.035\%$ relative shift from $\alpha{=}0$). These results confirm that our centralization findings do not hinge on the choice of $\alpha$. As shown in Table~\ref{tab:robustness}, age- and recency-adjusted variants reduce inequality by 7–14\% while preserving high rank stability.

We now project the benchmark ecosystem onto a tripartite graph whose nodes represent \emph{benchmarks}, \emph{authors}, and \emph{institutions} (Table \ref{labelnetwork}), with institution nodes inferred from paper-year author metadata. In the latest snapshot this graph comprises $2{,}402$ nodes connected by $4{,}559$ undirected edges. To characterise network structure we measure (i) degree centrality $d(v)=\deg(v)/(N-1)$ for every node, (ii) the Gini coefficient of the degree-centrality distribution, $G=0.477$, and (iii) betweenness centrality on the largest 3-core (390 nodes). Degree reveals hubs: the top-ranked benchmark alone links to $18.8\%$ of all actors, while the first-, second-, and third-ranked institutions record institutional degrees of $0.175$, $0.062$, and $0.052$, respectively. Betweenness highlights bridges: the two highest-betweenness authors each carry more than $3.5\%$ of all shortest paths in the 3-core, indicating pivotal roles in connecting otherwise disjoint author clusters \cite{Kito2024Litigation,Broido2019ScaleFree}.

This dual dynamic—rapid expansion of the benchmark community on one hand, and concentrated evaluative influence on the other—highlights a structural pattern in the AI model ecosystem. Even as the barrier to creating new benchmarks has lowered and participation has widened, the benchmarks that shape de facto standards and garner the most attention tend to emerge from a concentrated group of contributors. In effect, the community’s sense of ``what matters'' in evaluating AI is co-defined by many voices, with central actors helping to provide shared yardsticks and shape focus via path dependence.

\begin{table}[t]
\centering
\setlength{\tabcolsep}{3.5pt}
\renewcommand{\arraystretch}{1.1}
\caption{Robustness of benchmark authority to age/recency adjustment. Jaccard uses $k\!=\!10$.}
\label{tab:robustness}
\footnotesize
\begin{tabular}{lrrrrrr}
\toprule
\textbf{Variant} & \textbf{Gini} & \,$\Delta$\,\textbf{Gini} & \textbf{HHI} & \,$\Delta$\,\textbf{HHI} & $\boldsymbol{\rho}$ & \textbf{J10} \\
\midrule
Baseline (cumulative) & 0.675 & \phantom{+}0.0\%  & 0.020 & \phantom{+}0.0\%  & 1.000 & 1.000 \\
Rate/age ($\ge$0.25y) & 0.578 & $-14.4$\% & 0.015 & $-25.0$\% & 0.899 & 0.538 \\
Window 1y             & 0.585 & $-13.3$\% & 0.015 & $-25.0$\% & 0.928 & 0.538 \\
Window 2y             & 0.601 & $-11.0$\% & 0.016 & $-20.0$\% & 0.955 & 0.538 \\
Window 3y             & 0.623 & \;\,$-7.7$\% & 0.017 & $-15.0$\% & 0.979 & 0.538 \\
Decay $h{=}1$y        & 0.604 & $-10.5$\% & 0.016 & $-20.0$\% & 0.965 & 0.538 \\
Decay $h{=}2$y        & 0.624 & \;\,$-7.6$\% & 0.017 & $-15.0$\% & 0.987 & 0.667 \\
Decay $h{=}3$y        & 0.636 & \;\,$-5.8$\% & 0.017 & $-15.0$\% & 0.993 & 0.818 \\
Decay $h{=}5$y        & 0.648 & \;\,$-4.0$\% & 0.018 & $-10.0$\% & 0.997 & 0.818 \\
\bottomrule
\end{tabular}

\vspace{2pt}
\scriptsize \textit{Notes:} $\Delta$ values are relative to the baseline. $\,\rho$ is Spearman over the top-20 union. J10 is top-10 Jaccard vs.\ baseline (0.538 $\approx$ 7 shared, 0.667 $\approx$ 8, 0.818 $\approx$ 9).
\end{table}

\section{Trade-offs in benchmark concentration}
\label{sec:conc_eval_structures}

When a small set of benchmarks becomes widely adopted, evaluation provides shared yardsticks that reduce noise and aid comparability, while also shaping research focus via path dependence—what is measured tends to be optimized. Teams tune to leaderboards; a single widely used test by a central actor can steer architecture choices and training signals, as ImageNet did for vision and GLUE$\rightarrow$SuperGLUE for NLP \cite{Russakovsky2015ImageNet,Wang2018GLUE,Wang2019SuperGLUE}. In this way, concentration can both simplify coordination and reinforce path dependence, potentially leaving capability areas that are not directly rewarded with less attention.

Stewarded benchmarks also exhibit path dependence toward central actors. Even when code and data are open, the stewarding institution coordinates which tasks are added, how scores are computed, and what counts as failure. Update cycles reflect legitimate priorities and resource constraints, which can shift metric emphasis and incidentally favor familiar architectures or tooling, even absent explicit coordination \cite{singh2025leaderboard}. The resulting agenda mirrors measured influence and investment patterns, offering a coherent reference point while entailing opportunity costs for unmeasured directions.

A third consideration is information salience. Investors, policymakers, and journalists increasingly use benchmark scores as shorthand for “how good” systems are. With only a few highly visible tests, over-optimization to narrow task sets can overstate general capability and contribute to boom–bust dynamics in expectations. By contrast, a portfolio of complementary benchmarks makes narratives more robust by offering multiple lenses on safety and utility.

In sum, concentrated evaluative structures offer coordination benefits and shared yardsticks, while also introducing trade-offs: they can amplify over-optimization incentives, make it harder for novel ideas or failure cases to surface quickly, and allow narratives to be shaped disproportionately by a small set of visible tests.

\section{Agent-based simulation of coordination and concentration dynamics}
\label{sec:simulation}

The previous section documented that concentrated benchmark regimes can provide shared yardsticks and economies of scale in evaluation, alongside familiar trade-offs such as path dependence. The natural next question is \emph{what concrete forces push the ecosystem toward high concentration or, alternatively, sustain diversity while preserving coordination benefits?} To obtain a first, mechanism-level answer we build a deliberately stripped‐down agent-based model that retains the three behaviours most frequently cited in empirical work: attraction to popular leader-boards, fatigue with overfit tests, and the occasional birth of entirely new benchmarks.

Formally, time advances in discrete steps.  
At each step a new AI evaluator arrives and, with probability~\(\gamma\),
publishes a fresh benchmark; otherwise she chooses an incumbent
\(B_i\) with probability
\[
P_i(t)=
\frac{\bigl[A_i(t)\bigr]^{\alpha}\,
      \exp\!\bigl[-\beta\,O_i(t)\bigr]}
     {\sum_j
      \bigl[A_j(t)\bigr]^{\alpha}\,
      \exp\!\bigl[-\beta\,O_j(t)\bigr]},
\]
where \(A_i(t)\) is the benchmark’s accumulated authority 
(citations, stars, leaderboard entries) and \(O_i(t)\) is an
“over-fit debt’’ that increments whenever the same test is reused.
After selection we set \(A_i\!\leftarrow\!A_i+1\) and
\(O_i\!\leftarrow\!O_i+1\); all other debts decay by a small constant
\(\delta\), modelling eventual forgiveness of staleness.
The exponent \(\alpha>1\) captures the well-documented
Matthew effect whereby popular artefacts attract yet more attention,
while \(\beta>0\) measures how strongly communities shy away from tests
perceived as gamed.

\begin{figure*}[!t]
  \centering
  \includegraphics[width=\linewidth]{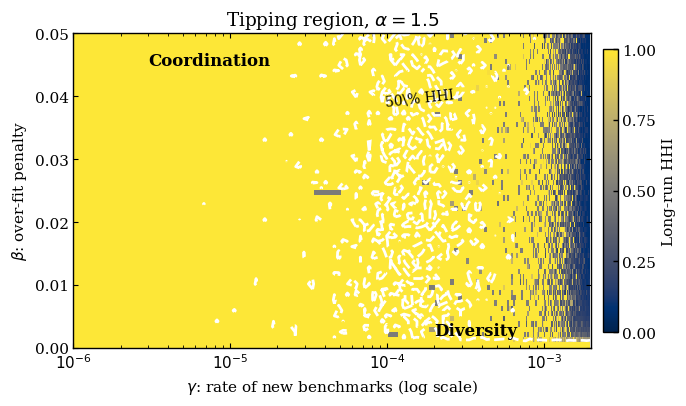}
  \caption{Steady‐state benchmark concentration as a function of the
           over-fit penalty~\(\beta\) and the birth rate of new
           benchmarks~\(\gamma\) (log scale, \(\alpha=1.5,\ \delta=0.1\)).
           The dashed contour shows the \(\mathrm{HHI}=0.5\) tipping line.}
  \label{fig:tipping}
\end{figure*}

We run the simulation for \(N=10^4\) steps over a grid of
\(\beta\in[0,0.05]\) and \(\gamma\in[10^{-6},2\times10^{-3}]\),
holding \(\alpha=1.5\) and \(\delta=0.1\).
Figure~\ref{fig:tipping} plots the resulting steady-state concentration using
the Herfindahl–Hirschman Index
\(\mathrm{HHI}=\sum_i (A_i/\!\sum_j A_j)^2\).
Bright yellow indicates high concentration (\(\mathrm{HHI}\approx1\)),
deep blue indicates a pluralistic field
(\(\mathrm{HHI}\approx0\)); the white dashed line marks the locus
\(\mathrm{HHI}=0.5\).

Three features stand out.  
First, when the influx of fresh benchmarks is essentially zero
(\(\gamma\to0\)) the system is pulled into a single central actor
regardless of how harshly we penalise over-fitting; preferential
attachment dominates.  
Second, once even a faint trickle of new tests appears
(\(\gamma\gtrsim10^{-4}\) in our units) concentration collapses:
\(\mathrm{HHI}\) dives below 0.2 and remains low almost
independently of~\(\beta\).
Third, increasing the over-fit penalty shifts the tipping line only
marginally; the decisive control variable is the
\emph{creation rate of novel benchmarks}, not the severity of the
penalty imposed on stale ones.

In our model, higher entry rates of new benchmarks are associated with lower steady-state concentration, whereas penalizing re-use has a smaller effect.

\section{Discussion}

Our findings reveal a rapidly evolving AI model ecosystem where growth in scale and participation brings coordination needs. On the model-supply side, the rise in foundation models and the diversification of their sources signal a broadening of development; at the same time, accompanying declines in documentation and accessibility can raise sensemaking and governance demands. On the evaluation side, we observe complementary dynamics: an expansion in the number and variety of benchmarks (and the researchers creating them) alongside concentrated benchmark influence among central actors, which can provide shared yardsticks while carrying familiar trade-offs. 

Point estimates hint at a modest decline in concentration—roughly 14\% per year for benchmarks and 28\% for models—but the 95\% confidence intervals include zero ($p \approx 0.1$--$0.3$). Accordingly, we do not reject a null of no change at conventional levels: centralization has remained broadly stable over the sample period. Our coverage checks also clarify what our authority metric captures: contemporaneous evaluative \emph{salience} rather than curated coverage. When we compare our top lists to external registries, overlap is limited but interpretable. Against the HELM core scenarios the Jaccard index is $\approx 0.05$ for the top–10 and $\approx 0.11$ for the top–20, with three shared items (MMLU, GSM8K, MATH) and a moderate rank correlation on the intersection ($\rho \approx 0.50$). The original Open-LLM leaderboard tasks share two items (MMLU, GSM8K; Jaccard $\approx 0.08$; $\rho$ not informative with two items), while the updated Open-LLM v2 set and the Swallow v2 English set show no overlap in our snapshot. This pattern is consistent with different design goals: coverage lists emphasize methodological breadth and stability, whereas our authority index reflects where evaluative attention is currently concentrated in practice, including dialogue, coding, agentic behaviour, and safety platforms that have surged since 2023.

These dynamics raise several implications for the AI research community and policymakers \cite{frank2019evolution, cebrian2025supervision}.
First, the declining transparency and accessibility of model releases may increase information frictions.
In our data, documentation quality and open access have not kept pace with the boom in model development
(see Fig.\ref{fig:access}).
If this pattern persists, information asymmetries could grow: some organizations may retain fuller knowledge of cutting-edge models’ capabilities and training details, while others (including regulators and smaller labs) have sparser information.
Strengthening reporting standards—\emph{e.g.}, comprehensive model cards \cite{mitchell2019modelcards}—and, where appropriate, incentivizing open-weight releases may support comparability, reproducibility, and reuse \cite{NatureEditorial_2025_LLM_PeerReview,Gibney_2025_DeepSeek_Landmark_Paper}.

\begin{table*}[!ht]
    \centering
    \caption{Summary of network structure and top‐central entities in the
             benchmark--author--institution graph 
             ($N_{\text{nodes}}=2{,}402$, $E_{\text{edges}}=4{,}559$, 
             degree‐centrality Gini $=0.477$;
             betweenness computed on the $k{=}3$ core, $N=390$).}
    \label{tab:network_summary}
    \small
    \begin{tabular}{@{}llc@{}}
        \toprule
        \multicolumn{3}{c}{\textbf{Top 10 benchmark hubs (by degree centrality)}}\\
        \midrule
         1 & BigBench                       & 0.188 \\
         2 & TrustLLM                       & 0.029 \\
         3 & HumanEval                      & 0.024 \\
         4 & HELM                           & 0.021 \\
         5 & LegalBench                     & 0.017 \\
         6 & AnthropicRedTeam               & 0.015 \\
         7 & FOLIO                          & 0.015 \\
         8 & BigCodeBench                   & 0.014 \\
         9 & BiGGen‐Bench                   & 0.013 \\
        10 & HHH (Helpfulness, Honesty, Harmlessness) & 0.013 \\
        \bottomrule
    \end{tabular}
    \hfill
    \begin{tabular}{@{}llc@{}}
        \toprule
        \multicolumn{3}{c}{\textbf{Top 10 author hubs (by degree centrality)}}\\
        \midrule
         1 & Dan Hendrycks                 & 0.007 \\
         2 & Yejin Choi                    & 0.007 \\
         3 & Samuel R.~Bowman              & 0.006 \\
         4 & Percy Liang                   & 0.005 \\
         5 & Dawn Song                     & 0.005 \\
         6 & Collin Burns                  & 0.005 \\
         7 & Mantas Mazeika                & 0.004 \\
         8 & Andy Zou                      & 0.004 \\
         9 & Steven Basart                 & 0.004 \\
        10 & Bo Li                         & 0.004 \\
        \bottomrule
    \end{tabular}

    \vspace{1em}

    \begin{tabular}{@{}llc@{}}
        \toprule
        \multicolumn{3}{c}{\textbf{Top 10 institutions (by degree centrality)}}\\
        \midrule
         1 & Google                         & 0.175 \\
         2 & Microsoft                      & 0.062 \\
         3 & Stanford University            & 0.052 \\
         4 & Carnegie Mellon University     & 0.040 \\
         5 & UC Berkeley                   & 0.037 \\
         6 & Tsinghua University            & 0.035 \\
         7 & \textit{Unknown / unlisted}    & 0.032 \\
         8 & University of Washington       & 0.027 \\
         9 & Peking University              & 0.024 \\
        10 & OpenAI                         & 0.015 \\
        \bottomrule
    \end{tabular}
    \hfill
    \begin{tabular}{@{}llc@{}}
        \toprule
        \multicolumn{3}{c}{\textbf{Top 10 authors (betweenness in $k=3$ core)}}\\
        \midrule
         1 & Yejin Choi                    & 0.040 \\
         2 & Jie Tang                      & 0.035 \\
         3 & Yixin Liu                     & 0.026 \\
         4 & Bo Li                         & 0.021 \\
         5 & Dan Hendrycks                 & 0.021 \\
         6 & Ion Stoica                    & 0.020 \\
         7 & Mantas Mazeika                & 0.018 \\
         8 & Percy Liang                   & 0.017 \\
         9 & Graham Neubig                 & 0.017 \\
        10 & Nikita Nangia                 & 0.017 \\
        \bottomrule
    \end{tabular}
    \label{labelnetwork}
\end{table*}

Second, the persistence of concentrated benchmark influence invites reflection on how evaluative standards are set in AI. When a small cluster of actors disproportionately shapes the benchmarks that define success (e.g., popular leaderboards or canonical test sets), there is the possibility of narrower evaluative lenses \cite{avin2021filling, mitchell2019modelcards, brundage2020toward, mokander2023auditing, Chaqfeh2023DigitalInequality}. Certain tasks or values may be emphasized while others receive less attention. Such concentration can incidentally underweight some perspectives; for instance, benchmarks originating predominantly from English-speaking or Western institutions may underrepresent challenges pertinent to other languages, cultures, or policy contexts. Encouraging wider participation—including international and historically underrepresented research communities—in developing and critiquing benchmarks can help diversify the evaluative toolkit \cite{HernandezOrallo2014Universal}. In this light, programs that fund collaborative benchmark development across institutions or that support “benchmark audits” (analogous to model audits) may further broaden coverage \cite{zhou2025general,weidinger2025toward}.

Third, our analysis underscores the coupling between the model and benchmark layers of the LLM ecosystem.
Influence over evaluation follows a classic heavy-tailed pattern: a few organizations concentrate measured “benchmark authority” while a long tail remains marginal \cite{Price1965Networks,Merton1968Matthew,Clauset2009Power}. Such inequality is consistent with preferential-attachment models of network growth, wherein early or well-resourced actors attract disproportionate citations and reuse \cite{Barabasi1999Emergence}.
In practice, a lab that launches a widely adopted benchmark can rapidly accrue further attention, reinforcing measured influence via self-reinforcing dynamics \cite{Salganik2006Inequality,Clauset2008Hierarchy}.
Some centralization can be beneficial—shared reference suites ease comparability—while high concentration may entail entry and over-optimization trade-offs.

Finally, our results speak to ongoing policy discussions at national and international levels. Evaluation has become a recurring theme in proposals for AI oversight—for example, the EU AI Act includes provisions related to transparency and risk management for certain AI systems \cite{euai2023}. A pluralistic benchmark landscape can inform these debates by offering diverse measures of risk (e.g., robustness, bias, environmental impact \cite{luccioni2024environmental}) and by helping to substantiate claims about system performance \cite{brundage2020toward}. At the same time, when evaluative influence is concentrated, the picture of capabilities and risks that reaches decision makers may be narrower. Broad participation in the development of assessment standards—for instance, through interdisciplinary committees or international bodies—could help align metrics with wider expertise rather than the priorities of a small contributor set \cite{cui2024risk, anwar2024foundational, weidinger2021ethical, prunkl2021institutionalizing}.

\section{Limitations}

Our findings rest on two public snapshots—the Stanford Ecosystem Graph and the Evidently AI benchmark registry—that inevitably omit models and benchmarks released outside those time windows or never indexed at all. 
Because our inclusion criteria (public paper + code + data under a permissive license) favor well-documented, English-first projects, our benchmark slice may under-represent community or regional practices. In the eligible subset (\(n=134\)), only \(\mathbf{13}\) suites explicitly target fairness/bias (\(\sim 10\%\)) and \(\mathbf{4}\) target toxicity (\(\sim 3\%\)), indicating topical skew. At the same time, model production is multipolar— with substantial activity from Chinese and U.S. organizations—so any centralization we report should be read as the structure of the openly documented benchmark layer circa June 2025, not the full space of global evaluation practice.

Even within the captured records, metadata completeness varies significantly. Approximately one-quarter of benchmarks lack structured author–affiliation information, necessitating heuristic inference of institutions from email domains or leaving affiliations as \textit{unknown}. Such inferred affiliations introduce uncertainty, as misclassifications could meaningfully influence institutional concentration metrics. Furthermore, citation counts sourced from Semantic Scholar might miss GitHub-only releases or double-count successive arXiv versions. GitHub \textit{stars}, employed as proxies for developer engagement, can similarly be influenced by promotion or transient attention cycles. These noisy signals propagate directly into our authority measure and centrality rankings, which should therefore be read as indicative rather than definitive.

Institutional naming ambiguity introduces additional uncertainties. Although alias tables reconcile obvious institutional name variants, they cannot exhaustively disambiguate subsidiaries, minor variations, or overlapping author identities (e.g., frequent names). Our method of assigning equal fractional credit when benchmarks have multiple institutional co-authors further simplifies complex, often asymmetrical contributions—such as lead institutions providing primary datasets while collaborators supply minor validation roles—so our institutional shares reflect measured co-authorship rather than a full accounting of contribution intensity.

Moreover, specific methodological choices in network construction influence our structural insights. Our decision to compute centrality metrics primarily on the network's $k!=!3$ core may understate the significance of peripheral contributors or emerging actors, potentially obscuring meaningful innovation occurring outside core structures. The weighting of benchmark influence—calculated as $\log(1+\text{citations}) + 0.25,\log(1+\text{stars})$—also introduces arbitrariness, as alternative weighting schemes (different values of $k$, weighted edges, or time-normalized citation rates) could yield different centralization estimates.

Collectively, these limitations underscore that our measures of evaluative concentration should be viewed as indicative rather than definitive; missing data, uncertain affiliations, methodological simplifications, and implicit scope assumptions may moderate or amplify the centralization we report. At the same time, concentrated benchmarks can provide shared yardsticks that help organize evaluation amid rising heterogeneity, especially within the curated scope of our datasets.

\section{Conclusion}

In our 2025 snapshot, the LLM ecosystem is expanding rapidly and becoming more heterogeneous, with model creation dispersing even as benchmark influence exhibits a heavy-tailed, concentrated pattern. This concentration can provide coordination benefits—shared yardsticks that support standardization, comparability, and reproducibility—while posing familiar, bounded trade-offs (e.g., path dependence and over-optimization). Our network analysis documents where measured (citation- and usage-based) influence concentrates across benchmarks, authors, and institutions; in a simple agent-based simulation, higher rates of benchmark entry are associated with lower steady-state concentration, whereas stronger penalties for re-use have comparatively smaller effects.

Taken together, these results point to a balanced path: widely recognized reference suites can help stabilize evaluation amid complexity, while a broader portfolio of well-documented, auditable benchmarks can enrich coverage across tasks, languages, and modalities. Within the limits of our curated datasets and observational design, these structural patterns offer a coherent lens for sensemaking in a fast-moving field and may provide a practical basis for aligning evaluation with emerging capabilities.

\section*{Data and Code Availability}

All analysis scripts, derived datasets, and figure-generation notebooks are available at \url{https://github.com/manuelcebrianramos/llm-benchmark-topology}. The repository fully reproduces all results and figures using only two openly licensed sources: 

\begin{itemize}
\item Stanford Foundation-Model Ecosystem Graph (snapshot: March 1, 2025; license: CC-BY 4.0). Available at: \url{[https://crfm.stanford.edu/ecosystem/}]

\item Evidently AI LLM Benchmark Registry (snapshot: June 12, 2025; license: Apache 2.0). Available at: \url{[https://www.evidentlyai.com/llm-evaluation-benchmarks-datasets}]
\end{itemize}

No proprietary data or closed-source software are required to replicate this study.

\vspace{1cm}

\section*{Acknowledgments}

We thank Lexin Zhou, David García and José Hernández-Orallo for discussions on this line of research. MC acknowledges support from grant PID2023-150271NB-C21 from the Spanish Ministry of Science, Innovation, and Universities (MICINN) / Spanish State Research Agency (AEI, DOI: 10.13039/501100011033). Additional funding was provided by Google.org through the Silicon Valley Community Foundation via a grant to the Fundación General CSIC. The funders had no role in the study design, data collection and analysis, decision to publish, or preparation of the manuscript.

\bibliographystyle{unsrt}
\bibliography{references}

\end{document}